# Monolithically integrated 940 nm half VCSELs on bulk Ge substrates

YUNLONG ZHAO,[1] JIA GUO,[1] WAN ZEYU,[1] MARKUS FEIFEL,[2] HAO-TIEN CHENG,[3] YUN-CHENG YANG,[4] LUKAS CHROSTOWSKI,[1] DAVID LACKNER,[2] CHAO-HSIN WU,[3,4] AND GUANGRUI (MAGGIE) XIA[1,*]

[1] *Department of Materials Engineering, the University of British Columbia, Vancouver BC, V6T 1Z4 Canada*
[2] *Fraunhofer Institute for Solar Energy Systems (ISE), Heidenhofstraße 2, 79110 Freiburg, Germany*
[3] *Graduate Institute of Electronics Engineering, National Taiwan University, Taiwan, ROC*
[4] *Graduate Institute of Photonics and Optoelectronics, National Taiwan University, Taiwan, ROC*
[5] *Department of Electrical and Computer Engineering, University of British Columbia, Vancouver, BC, V6T 1Z4 Canada*
[*]*gxia@mail.ubc.ca*

**Abstract:** High quality n-type AlGaAs distributed Bragg reflectors (DBRs) and lnGaAs multiple quantum wells were successfully monolithically grown on 4-inch off-cut Ge (100) wafers. The grown structures have photoluminescence spectra and reflectance spectra comparable to those grown on conventional bulk GaAs wafers and have smooth morphology and reasonable uniformity. These results strongly support full VCSEL growth and fabrication on larger-area bulk Ge substrates for the mass production of AlGaAs-based VCSELs.

## 1. Introduction

The demand for vertical cavity surface emitting lasers (VCSELs) has dramatically increased in past few years due to its applications as near infrared illumination sources in three-dimension (3D) sensing and imaging, crucial to some of the most popular features in smartphones, virtual reality and augmented reality applications, such as Face ID and proximity-sensing [1-2]. Compared with other laser technology, the benefits of VCSEL technology include scalable output power, high quality optical beam, high wall-plug efficiency, stable wavelength over temperature, low spectral width, easy testing and packaging [3]. VCSELs are also widely used in the light detection and ranging (LiDAR) system in electric vehicles and autonomous mobile robots. As a result, the global VCSEL market was predicted to increase further to $2.4 billion by 2026 from the $1.2 billion market size in 2021 at a 13.6% compound annual growth rate [4-7]. Up to now, VCSEL production still largely relies on 3-, 4- and 6-inch bulk GaAs wafers, which limits the production volume [8]. Scaling up the VCSEL production by using larger wafers is a very effective way to solve this problem, which is especially beneficial to boost larger VCSEL array production used in high power LiDAR.

The major concern in the epitaxy of AlGaAs VCSELs on GaAs substrates is the inherent strain formed during VCSEL growth. A typical VCSEL epitaxial structure is about 5 to 15 microns thick, and the majority of the thickness is in the bottom and the top distributed Bragg reflectors (DBRs) made of $Al_xGa_{1-x}As/Al_yGa_{1-y}As$ superlattices, x < y.

GaAs (the lower limit of $Al_xGa_{1-x}As$), Ge and AlAs (the higher limit of $Al_yGa_{1-y}As$), have lattice constants of 5.653, 5.658 and 5.660 Å, respectively. For the production on 3-inch and 4-inch GaAs wafers, the lattice mismatch induced strain, 0.14% lattice constant mismatch between AlAs and GaAs, at room temperature is acceptable [9]. When the growth is on larger GaAs wafers, this strain results in significant bow and warp on wafers after growth, which lead to low chip yield and reliability problems. A solution by replacing bulk GaAs substrates with bulk Ge substrates was proposed. Ge lattice constant is between that of $Al_xGa_{1-x}As$ and $Al_yGa_{1-y}As$, reducing the lattice mismatch strain and thus the bow and warp of the Ge substrates after the growth [10], which make larger diameter VCSEL substrates feasible for mass production. Ge wafers also have some other advantages including more mechanically robust, thinner thickness and less density of threading dislocations (TDD), which are crucial to lower the yield loss and reduce reliability failures in VCSELs on large size GaAs substrates [11]. Commercially, the price of 6" Ge wafers is not higher than 6" GaAs wafers. More importantly, Ge wafers can be made to 8- and 12-inch diameter, while GaAs wafers are too brittle to make for these sizes. Therefore, 6 to 12-inch diameter Ge wafers are promising solutions for the mass production of VCSELs without significant fabrication cost increase. In 2020, IQE first demonstrated 940 nm VCSELs on 6" bulk Ge wafer. With unoptimized processes for Ge-based VCSELs, they produced comparable DBR stopbands, active region photoluminescence (PL), lasing performance as GaAs-based VCSELs and have much less wafer bow and warp [12]. It is the only report on bulk Ge-based full VCSELs so far. However, no Ge substrate specifications, Ge to GaAs transition layers, growth and testing conditions were revealed in their presentation. Later, we achieved the successful monolithic integration of $Al_xGa_{1-x}As$ n-type distributed Bragg reflectors (n-DBRs) on bulk Ge substrates independently, with an InGaP nucleation layer, InGaAs, GaAs transition layers on bulk Ge [13]. SiN backside coating was needed to prevent Ge sublimation. The n-DBRs on GaAs/InGaAs/InGaP/bulk-Ge/SiN substrates have comparable stopbands and smooth surface morphology as bulk GaAs-based DBRs [13].

Since then, we continued to develop the epitaxy toward bulk Ge-based VCSELs. In this paper, we report the successful monolithic integration of $In_xGa_{1-x}As$ multiple quantum wells (MQW) and $Al_xGa_{1-x}As$ n-type distributed Bragg reflectors (referred as half VCSELs thereafter) on 4-inch Ge full wafers. These bulk Ge-based half VCSELs produced comparable photoluminescence and stopband features as the GaAs counterparts. Important material, processing and testing conditions are discussed in detail.

## 2. Experiment Design and Epitaxy growth

The schematic structure of the half VCSELs grown on Ge substrates is shown in Fig. 1. The Ge wafers in this study were n-type 375 μm thick 4-inch (100) Ge wafers with 6-degree miscut towards <111> provided by Umicore. This thickness is the same as that of the Ge substrates used in the n-DBRs growth in Ref. [13]. The miscut is to reduce antiphase-domains (APDs) of the subsequent GaAs layers on Ge and to promote step-flow growth. To prevent Ge sublimation during high temperature processes, 100 nm SiN thin films were deposited on the Ge wafer backsides.

To transition from Ge to GaAs, an InGaP nucleation layer, a 950 nm n-$Ga_{0.985}In_{0.015}As$ layer with the lattice constant matched to Ge, and a 50 nm n-GaAs layer with $5\times10^{18}$ cm$^{-3}$ doping were grown on the Ge wafers. The top GaAs layer was expected to be lattice-matched to Ge. The growth was performed in an Aixtron 2800G4R metal organic chemical vapor deposition (MOCVD) reactor. After the backside SiN deposition and frontside GaAs/$In_{0.015}Ga_{0.985}As$/InGaP layer growth, the thickness of the 4-inch GaAs/InGaAs/InGaP/bulk-Ge/SiN wafers ready for DBRs and MQW epitaxy were 376 μm in total. A 2-degree offcut bulk 629 μm thick GaAs wafer provided by LandMark was used as the substrate of the control sample.

On the 4-inch GaAs/InGaAs/InGaP/bulk-Ge/SiN wafers and control GaAs wafer, a 500 nm thick GaAs layer was grown to have a cleaner surface for the subsequent n-DBRs. In principle, this base layer, n-DBR and active layer design should be updated according to the Ge substrate lattice constant. However, as this project's goal was to demonstrate lasing of VCSELs on Ge wafers and previous results using a GaAs base layer and n-DBR were promising, we decided to continue using a mature VCSEL design calibrated to GaAs substrates and will leave the design of VCSEL structure and composition for Ge wafers to future work.

Each DBR consists of 40 periods of n-type doped four layers: 20 nm $Al_xGa_{1-x}As$ (x: 0.12→0.9) with $3\times10^{18}$ $cm^{-3}$ doping / 56 nm $Al_{0.9}Ga_{0.1}As$ with $3\times10^{18}$ $cm^{-3}$ doping / 20 nm $Al_xGa_{1-x}As$ (x: 0.9→0.12) with $2\times10^{18}$ $cm^{-3}$ doping / 49 nm $Al_{0.12}Ga_{0.88}As$ with $2\times10^{18}$ $cm^{-3}$ doping (from the bottom to the top in one period). On top of the DBR, a half period of DBR consisting of 20 nm $Al_xGa_{1-x}As$ (x: 0.12→0.9) with $2\times10^{18}$ $cm^{-3}$ doping / 30 nm $Al_{0.85}Ga_{0.15}As$ with $2\times10^{18}$ $cm^{-3}$ doping was grown before the MQW epitaxy. The MQWs consist of three InGaAs quantum wells with GaAsP barriers. On top of the MQWs, one 100 nm undoped $Al_xGa_{1-x}As$ (x: 0.85→0.3) / 10 nm undoped GaAs are grown as the capping layers. The growth temperature was set to 760 °C and the vapor pressure was 50 mbar. The sources used for GaAs growth was $H_2$, $SiH_4$, $AsH_3$ and $Ga(CH_3)_3$ (TMGa). During the $Al_xGa_{1-x}As$ growth for bottom DBR layers, $Al_2(CH_3)_6$ (TMAl) was added to provide the Al source. For the MQW growth process, the sources to add In and P was $In(CH_3)_3$ (TMIn) and $PH_3$.

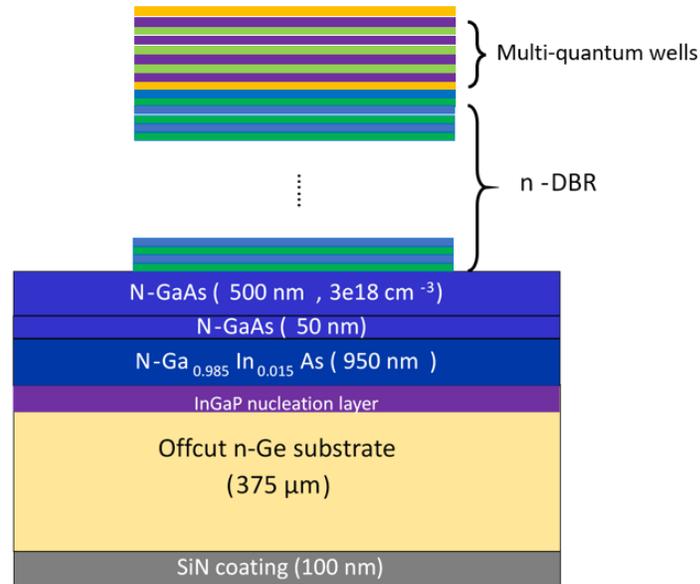

Fig. 1. Schematic structure of the half VCSELs grown on GaAs/InGaAs/InGaP/Ge/SiN substrate.

## 3. Results and discussions

To investigate the half VCSELs on the 4-inch GaAs/InGaAs/InGaP/bulk-Ge/SiN wafers, atomic force microscope (AFM) imaging, optical reflectance spectra measurements, scanning electron microscopy (SEM) were conducted in the material analysis. Photoluminescence spectroscopy (PL) was measured to characterize the MQW and the flatness of the full 4-inch wafers after the growth was checked by wafer bow and wrap measurement. Secondary ion mass spectrometry (SIMS) was performed to compare the chemical compositions of the QWs and the barrier layers.

*3.1 Surface quality*

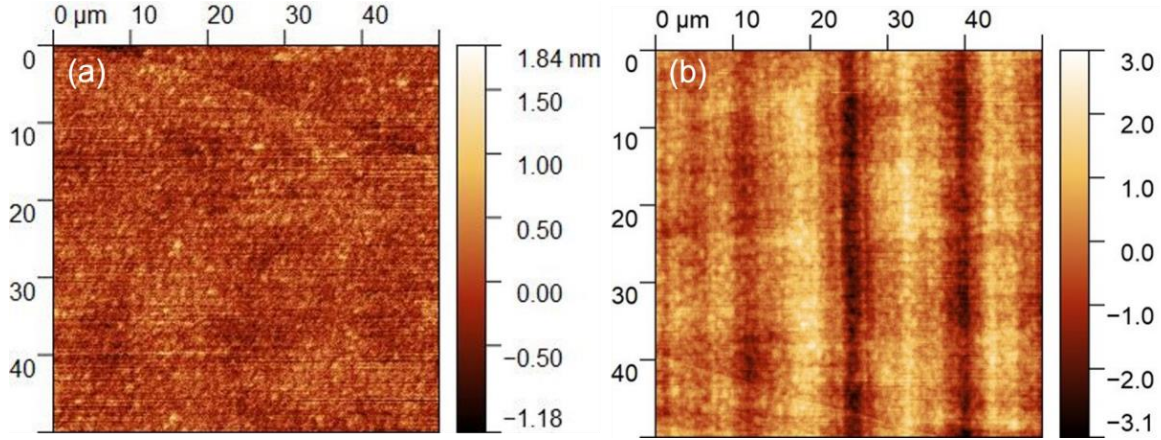

Fig. 2. AFM images of the surfaces of grown half VCSEL on (a) control GaAs wafer and (b) Ge wafer.

Good surface quality was confirmed under the optical microscope and AFM (Fig. 2) for the GaAs-based and Ge-based structures after MQW and DBRs growth. No antiphase domains or cracks were found. In Fig. 2(b), some minor crosshatch patterns can be seen in the AFM image of the Ge-based structure surface, which resulted from the small lattice mismatch between AlGaAs, GaAs and Ge. The root mean square (RMS) roughness measured by AFM for GaAs-based and Ge-based half VCSEL are 0.28 and 0.84 nm respectively measured on 50 µm x 50 µm areas, similar to the RMS roughness of the GaAs-based DBRs and Ge-based DBRs reported in [13]. The RMS roughness of the GaAs/InGaAs/InGaP/bulk-Ge/SiN substrates is 0.43 nm before the DBR growth.

*3.2 Reflectance*

The normal-incidence reflectance spectra of the GaAs-based half VCSEL and Ge-based half VCSEL were shown in Fig. 3. The reflectance spectra were measured with a Filmmetrix F20 thin-film analyzer, the same equipment used in the study of DBRs growth on Ge substrate [13]. All the reflectance values were normalized based on the peak reflectance value of the control sample GaAs-based half VCSEL. Fig. 3a is the comparison between the GaAs-based half VCSEL and the Ge-based half VCSEL. Both measurements were done at the center of the wafer. It shows that the Ge-based half VCSEL structures have comparable stopband shapes, widths and maximum peak heights as those of the control GaAs-based half VCSEL structures. The peak reflectance values of the Ge-based half VCSEL are 99.97% after normalization by the peak GaAs-based half VCSEL reflectance. The dips on the stopband result due to the absorption of the active region. At room temperature, the dip of the stopband of the GaAs-based half VCSEL locates at 925.7 nm wavelength and the counterpart from the Ge-based half VCSEL locates at 915.6 nm. The stopband dip positions are consistent with the PL peak wavelength discussed in 3.3, as they both reflect the energy band gaps of the MQWs.

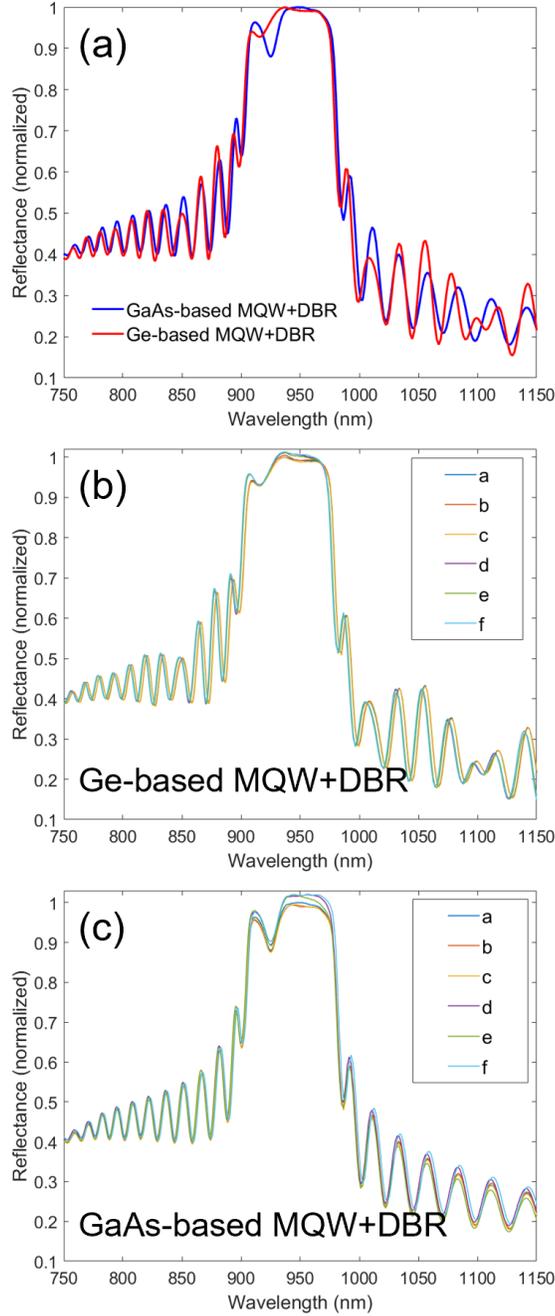

Fig. 3. (a) Normal-incidence reflectance spectra of the GaAs-based half VCSEL and the Ge-based half VCSEL. All spectra were taken at the center of the wafer and normalized by the maximum GaAs-based half VCSEL reflectance value. (b) Normal-incidence reflectance spectra of 6 locations of the Ge-based half VCSEL (a to f) placed from the central area to the edge of the Ge-based half VCSEL to show the cross-wafer growth uniformity. The reflectance values are normalized by the maximum GaAs-based half VCSEL reflectance value. (c) Normal-incidence reflectance spectra of 6 locations of the control GaAs-based half VCSEL (a to f) placed at different positions of the GaAs-based half VCSEL showing the cross-wafer growth uniformity, which are normalized by the maximum GaAs-based half VCSEL reflectance value.

Good reflectance spectra uniformity across the 4-inch Ge-based half VCSEL is observed as shown in Fig. 3(b). The spectra a was measured at the center of the 4-inch wafer. The spectra b – f is across the wafer from central area to the edge area to check the uniformity of the stopbands. The stopbands position shift across the wafer is less than 3.5 nm and the peak reflectance values are 99.88% to 101.23%. The small shift of stopbands position and fluctuation of peak reflectance values are expected due to flatness variation across the 4-inch wafer. The bow and warp of the Ge-based half VCSEL after growth measured by Tropel C100 system shows that there is a 90 µm height variation across the full wafer, which contributes to the non-uniformity of the stopbands. Fig. 3(c) shows the cross-wafer reflectance spectra of the control GaAs-based half VCSEL structures. The center of stopbands is at 939 nm to 940 nm, and the peak reflectance is 99.5% to 102%. The stopbands uniformity of Ge-based half VCSEL is comparable to its GaAs counterpart.

*3.3 PL measurements*

Photoluminescence spectroscopy is a non-contact, nondestructive method to characterize the light emission spectrum from the MQWs. The PL system used is equipped with a 532 nm continuous-wave (CW) laser and a liquid nitrogen Ge PIN diode detector made by Bruker Ltd. Fig. 4 shows the PL spectra of the Ge- and GaAs-based half

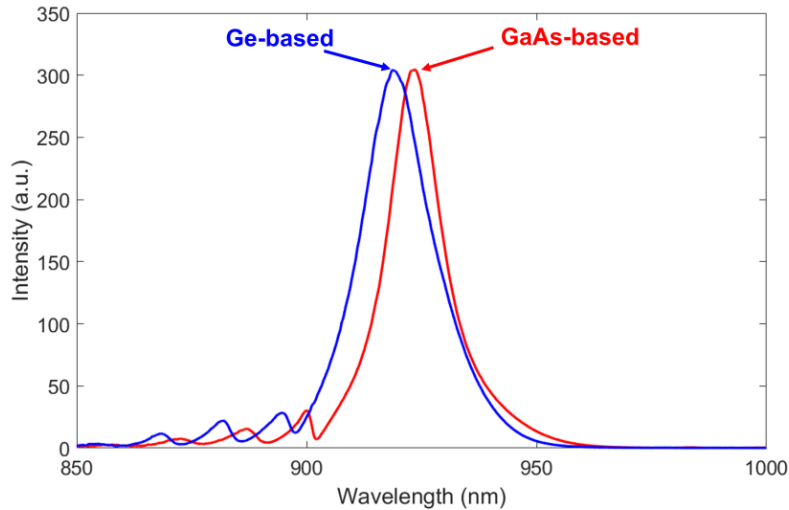

Fig. 4. Photoluminescence spectra of the GaAs-based half VCSEL and the Ge-based half VCSEL.

VCSEL. Both spectra share comparable peak intensity with a difference < 0.2%. The PL peak positions of the Ge- and GaAs-based half VCSEL are at 918.865 and 923.419 nm respectively. The 4.55 nm peak wavelength difference translates to 6.66 eV difference in the corresponding photon energy. This may be caused by the composition and strain difference of the MQWs, which were then investigated by SEM and SIMS discussed below. For the PL uniformity across the full wafer, the Ge-based half VCSEL shows an 8.54% intensity non-uniformity and a 0.028% peak wavelength non-uniformity, while the GaAs-based half VCSEL shows a 1.04% intensity and 0.014% peak wavelength non-uniformity.

*3.4 Cross-sectional SEM images*

The half VCSEL cross-sections were observed by SEM. The SEM samples were prepared by cleaving with a diamond scriber. Surface polishing or coating was not required. The images in Fig. 5 were collected using a FEI Nova NanoSEM system. The measurement mode was the immersion mode with a gaseous analytical detector. The operation voltage used was 10 kV and the working distance was 5.5 mm. No cracks were observed during measurement and the shadowy lines in Fig. 5(b) are due to the surface roughness of the sample, typical for a sample cleaved from an off-cut wafer.

Based on the cross-section SEM images of the half VCSELs in Fig. 5, good periodicity and layer uniformity of both types of structures can be confirmed. The thickness of MWQs was too thin to be measured accurately by SEM. The total thickness of Ge-based half VCSEL is 6.11 µm, which is in good agreement with the 6.10 µm thickness of the GaAs counterpart. This thickness different was not sufficient to explain the 6.66 eV difference in peak PL photon energy.

*3.5 SIMS measurement*

In order to check the composition of the MQWs, SIMS was performed to check the molar fractions of Al, Ga, As, P and In of the MQWs and barrier layers. The measurement was conducted by Eurofins EAG Laboratories. The depth scale was adjusted to match the SEM thickness. The accuracy of atom fractions is about +/- 5% in AlGaAs layer, and about +/- 20% in the mixed As/P layers.

Atomic fractions of In from the Ge- and GaAs-based MQWs are very close, and are within the SIMS accuracy, although SIMS profiling of the MQWs is heavily influenced by the SIMS intermixing effect. In the top and bottom GaAsP barrier layers, the Ge-based sample has around 3% more P than the GaAs-based sample, which corresponds to a bandgap difference of 0.039 eV and a 0.18% strain difference. This difference results in the dip wavelength shift in reflectance spectrum in Fig. 3(a) and the peak wavelength shift in PL spectrum in Fig. 4. Those wavelengths of Ge-based half VCSEL shift to shorter wavelength.

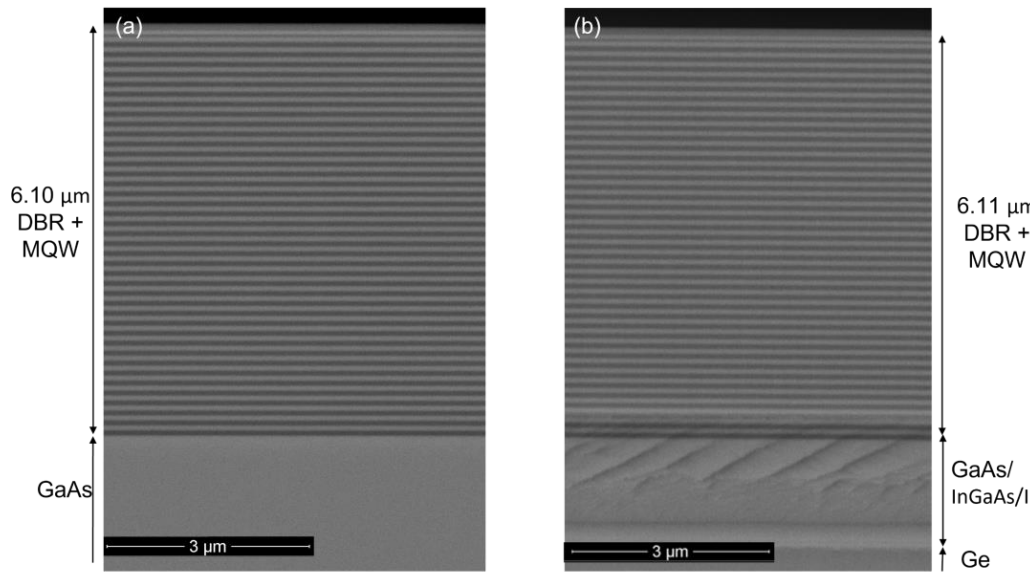

Fig. 5. Cross-section SEM images at 30K × magnifications of a (a) bulk GaAs-based half VCSEL, (b) Ge-based half VCSEL.

*3.6 Future work*

The results of half VCSEL growth are very encouraging. The next growth of full VCSEL epitaxy on 4-inch Ge wafers have been planned. Calibration runs to tune the Ge-VCSELs growth recipe to have the 940 nm lasering wavelength are on-going. More material analysis and device performance measurements will be conducted.

**4. Conclusion**

High quality half VCSELs with bottom n-DBR and MQWs were successfully grown monolithically on 4-inch Ge wafers. No APDs or cracks were formed in epitaxy on Ge wafers. The Ge-based half VCSELs have uniform reflectance spectra, PL peak intensity and PL peak wavelength comparable to those grown on the bulk GaAs wafer. Besides the first report of Ge-based VCSELs by IQE, this work independently developed bulk-Ge-based half VCSELs. Our results strongly support full VCSEL growth and fabrication on larger-area bulk Ge substrates for the mass production of AlGaAs-based VCSELs.

**5. Back matter**

**Funding.** Huawei Technologies, Canada (501100003816).

**Acknowledgments.** Umicore N. V., Belgium, is acknowledged for providing the bulk Ge wafers with GaAs/GaPAs epitaxy layers and SiN back coating.

**Disclosures.** The authors declare no conflicts of interest.

**Data availability.** Some data underlying the results presented in this paper are not publicly available at this time but may be obtained from the authors upon reasonable request.